\documentclass[prc,twocolumn,superscriptaddress,showpacs,twoside,floatfix]
{revtex4-1}

\usepackage{amssymb,epsfig}

\begin{document}

\title{Sensitivity of three-body decays to the reactions mechanism and the
initial structure by example of $^{6}$Be}

\author{L.V.~Grigorenko}
\affiliation{Flerov Laboratory of Nuclear Reactions, JINR, Dubna, RU-141980
Russia}
\affiliation{GSI Helmholtzzentrum f\"{u}r Schwerionenforschung, Planckstra{\ss}e
1, D-64291 Darmstadt, Germany}
\affiliation{National Research Center ``Kurchatov Institute'', Kurchatov sq.\ 1,
RU-123182 Moscow, Russia}
\author{I.A.~Egorova}
\affiliation{Bogoliubov Laboratory of Theoretical Physics, JINR, Dubna, 141980
Russia}
\author{R.J.~Charity}
\affiliation{Department of Chemistry, Washington University,
St.~Louis, Missouri 63130, USA.}
\author{M.V.~Zhukov}
\affiliation{Fundamental Physics, Chalmers University of Technology, S-41296
G\"{o}teborg, Sweden}

\date{\today. {\tt File: c:/latex/6be-s5/v9/6be-s5-9.tex }}

\begin{abstract}
The influence of the initial-state structure and the reaction mechanism on
three-body decays is investigated by example of the $^6$Be continuum populated
in neutron-knockout reactions on $^{7}$Be. The sensitivity of the $^{6}$Be
excitation spectrum and the three-body correlations to the different components
of the model
is demonstrated. It is shown that the spin composition of the initial state may
have an overwhelming effect on the three-body continuum. The characteristics and
structure of the second $0^+_2$ and $2^+_2$ states in $^{6}$Be are predicted
for the first time and the conditions for their reliable observations are
formulated. The effects of interference and the alignment of three-body states
on the three-body correlations are demonstrated.
\end{abstract}


\maketitle

%===============================================================================

\textit{Introduction.}
%
%===============================================================================
%
--- The study of systems beyond the nuclear driplines is an important trend of
modern low-energy nuclear research. Many of these systems belong to the
three-body or even few-body continuum and the reaction theory for populating
these states is not well developed. Modern high-precision experiments with
exotic beams require complicated analyses, advanced theoretical treatment, and
call for deeper insights in this field.

Observables in  reactions producing unbound systems depend on three major
ingredients: (i) the structure of initial nuclei, (ii) the reaction mechanism,
and (iii) the final-state interaction (FSI). For very narrow resonances
(extremely long-lived states), the aspects (i) and (ii) lose importance as the
structure formed in the reaction has enough time to ``forget'' how it was
created. Then for a consistent description of the system, it is sufficient to
study only the decay process (FSI effects) by itself. However, exactly when this
approach becomes valid is not always clear in advance. Clarity in this issue is
especially important for systems beyond the driplines, where the resonant states
(often already the ground states) are quite broad.

In this work, we demonstrate the importance of the reaction mechanism and the
initial-state structure for investigations of few-body systems beyond the
driplines. This is made by example of $^6$Be (three-body $\alpha+p+p$ continuum)
populated in neutron-knockout reactions from $^7$Be projectiles. The first
result of these studies has been published in Ref.~\cite{Egorova:2012}
elucidating the mechanism of democratic decay. Good agreement with experimental
data was demonstrated  for both the excitation spectrum and the three-body
energy-angular correlations over a broad range of excitation energy. However, in
this compact experimental work, many important theoretical issues of broader
interest by themselves, were left aside. In the present work, we would like to
focus on the most interesting theoretical results arising from our studies of
this data \cite{Egorova:2012}. This is a timely message considering the recent
high interest in the $^{6}$Be system
\cite{Grigorenko:2009,Grigorenko:2009c,Charity:2010,Papka:2010,Fomichev:2012}.
We also think that these results should have an important impact on our
understanding of three-body decays in general and aid in formulating
experimental strategies for studies of this phenomenon.

%-------------------------------------------------------------------------------
\begin{figure}
\includegraphics[width=0.48\textwidth]{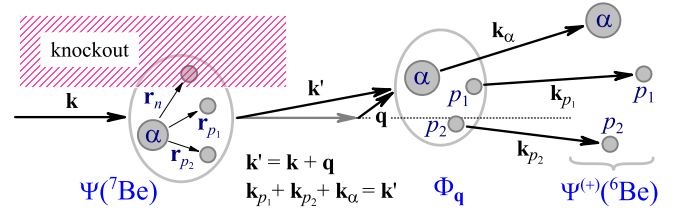}
\caption{Schematic illustrating the $^{6}$Be population formed
by  knocking a neutron out of a $^{7}$Be projectile.}
\label{fig:sketch}
\end{figure}
%-------------------------------------------------------------------------------

%===============================================================================

\textit{Theoretical model.}
%
%===============================================================================
%
--- The major features of the theoretical model were described in
Ref.~\cite{Egorova:2012}, but some details are needed in the context of
this work. The three-body $\alpha$+$p$+$p$ final-state interactions forming
the continuum of $^{6}$Be are described by solving the inhomogeneous three-body
Schr\"odingier equation
\begin{equation}
(\hat{H}_3 - E_T)\Psi^{(+)} = \Phi_{\mathbf{q}},
\label{eq:shred}
\end{equation}
where $\Psi^{(+)}$ is a wave function (WF) with pure outgoing asymptotics
obtained with the approximate boundary conditions of the three-body Coulomb
problem \cite{Grigorenko:2009,Grigorenko:2009c}. The energy $E_T$ is relative to
the $\alpha$+$p$+$p$ threshold. The source term $\Phi_{\mathbf{q}}$ depends on
the vector $\mathbf{q}$ of the transferred momentum and contains information
about the precursors and the reaction mechanism. The knockout of a neutron from
$^{7}$Be is described as a sudden neutron removal. After acting with the neutron
annihilation operator on the $^{7}$Be WF in coordinate space,  we obtain
\begin{equation}
\Phi_{\mathbf{q}} = \int d^3 r_n e^{i\mathbf{q r}_n} \langle
\Psi_{^4\text{\scriptsize
He}} | \Psi_{^7\text{\scriptsize Be}} \rangle .
\label{eq:sour}
\end{equation}
Here vector $\mathbf{r}_n$ points to the removed neutron, see Fig.\
\ref{fig:sketch}.

The $^{7}$Be WF is constructed as an ``inert'' $\alpha$-core plus $p$-wave
neutron and two protons with coupling
$[l_j(\nu)[l_j(\pi_1)l_j(\pi_2)]_J]_{J_{7\text{Be}}}$. The overlap integral with
the $\alpha$ particle for this WF is
\begin{eqnarray}
\langle \Psi_{^4\text{\scriptsize He}} | \Psi_{^7\text{\scriptsize Be}} \rangle
=  \alpha [p_{3/2} [p^2_{3/2}]_0]_{3/2} + \beta [p_{3/2} [p^2_{1/2}]_0]_{3/2}
\quad \nonumber \\
+  \gamma [p_{3/2} [p^2_{3/2}]_2]_{3/2} + \delta [p_{3/2}
[\frac{p_{3/2}p_{1/2}-p_{1/2}p_{3/2}}{\sqrt{2}}]_2]_{3/2} .\;\;\;
\label{eq:psi-7be}
\end{eqnarray}
After neutron removal, the terms with coefficients $\{\alpha, \beta\}$ lead to
the population of the $0^+$ state in $^{6}$Be while the terms with coefficients
$\{\gamma, \delta\}$ lead to the population of the $2^+$ state. Within the $0^+$
and $2^+$ configurations, the ratios $\alpha/\beta$ and $\gamma/\delta$ define
the spin composition of $\Phi_{\mathbf{q}}$, namely, the probability $W_S$ of
configurations with definite total spin $S$ (coupled spins of the two
``valence'' protons). The components of the source function with definite total
angular momentum $J$ can be written as
\begin{equation}
\Phi_{J,q} \sim \sqrt{1-\eta_J^2} \, |J, S=0 \rangle
+ \eta_J \,| J,S=1 \rangle ,
\label{eq:sour-spin}
\end{equation}
where the coefficients $\eta_J$, controlling the $W_{S=0}/W_{S=1}$ ratio, can be
expressed via $\{\alpha, \beta, \gamma, \delta\}$ using coefficients from Table
\ref{tab:recoupl}.

%All the above coefficients are real and vary at most in the range [-1,1].

%===============================================================================
\begin{table}[b]
\caption{The recoupling coefficients from shell-model-like $(jj)$ coupling in
the source to $(ls)$ coupling of the three-body model.}
\begin{ruledtabular}
\begin{tabular}[c]{ccccc}
$j_1, j_2, J:$  & $3/2,3/2,0$ & $1/2,1/2,0$ & $3/2,3/2,2$ & $3/2,1/2,2$   \\
\hline
$S=0$  &  $\sqrt{2/3}$   & $\sqrt{1/3}$  & $\sqrt{1/3}$  &  $\sqrt{2/3}$  \\
$S=1$  & $-\sqrt{1/3}$   & $\sqrt{2/3}$  & $\sqrt{2/3}$  &  $-\sqrt{1/3}$  \\
\end{tabular}
\end{ruledtabular}
\label{tab:recoupl}
\end{table}
%===============================================================================

For single-particle motion, we use the harmonic-oscillator WFs whose radial
behavior is
\begin{equation}
\phi(r) = (8/3\sqrt{\pi})^{1/2} (r^2/r_0^{5/2}) \exp[-r^2/(2r_0^2)].
\label{eq:osc}
\end{equation}
The value $r_0=1.8$ fm was used in the calculations which corresponds within our
model to the experimental matter radius of 2.31(2)~fm for $^{7}$Be.

%-------------------------------------------------------------------------------
\begin{figure}
\includegraphics[width=0.45\textwidth]{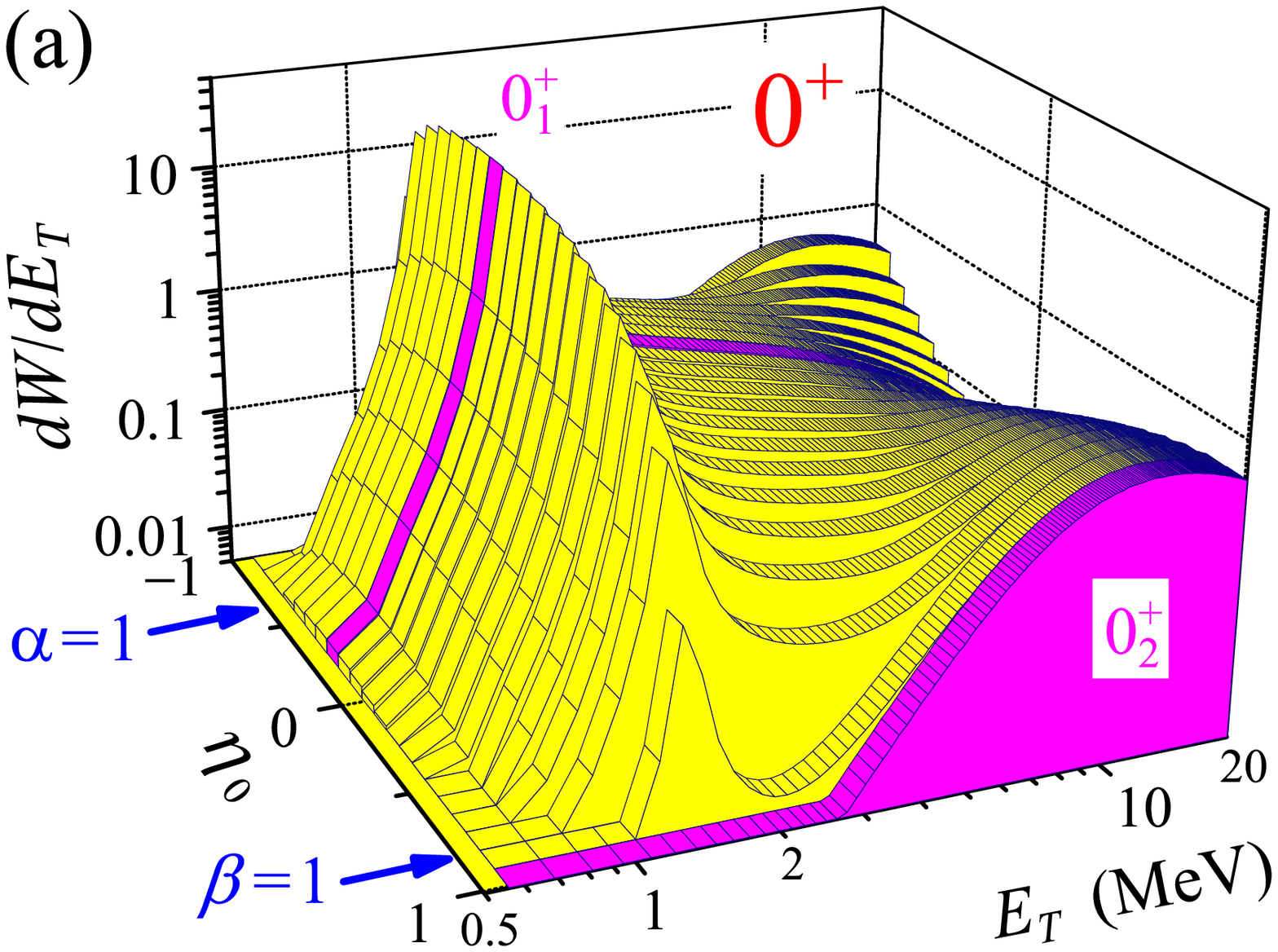} % 45
\includegraphics[width=0.44\textwidth]{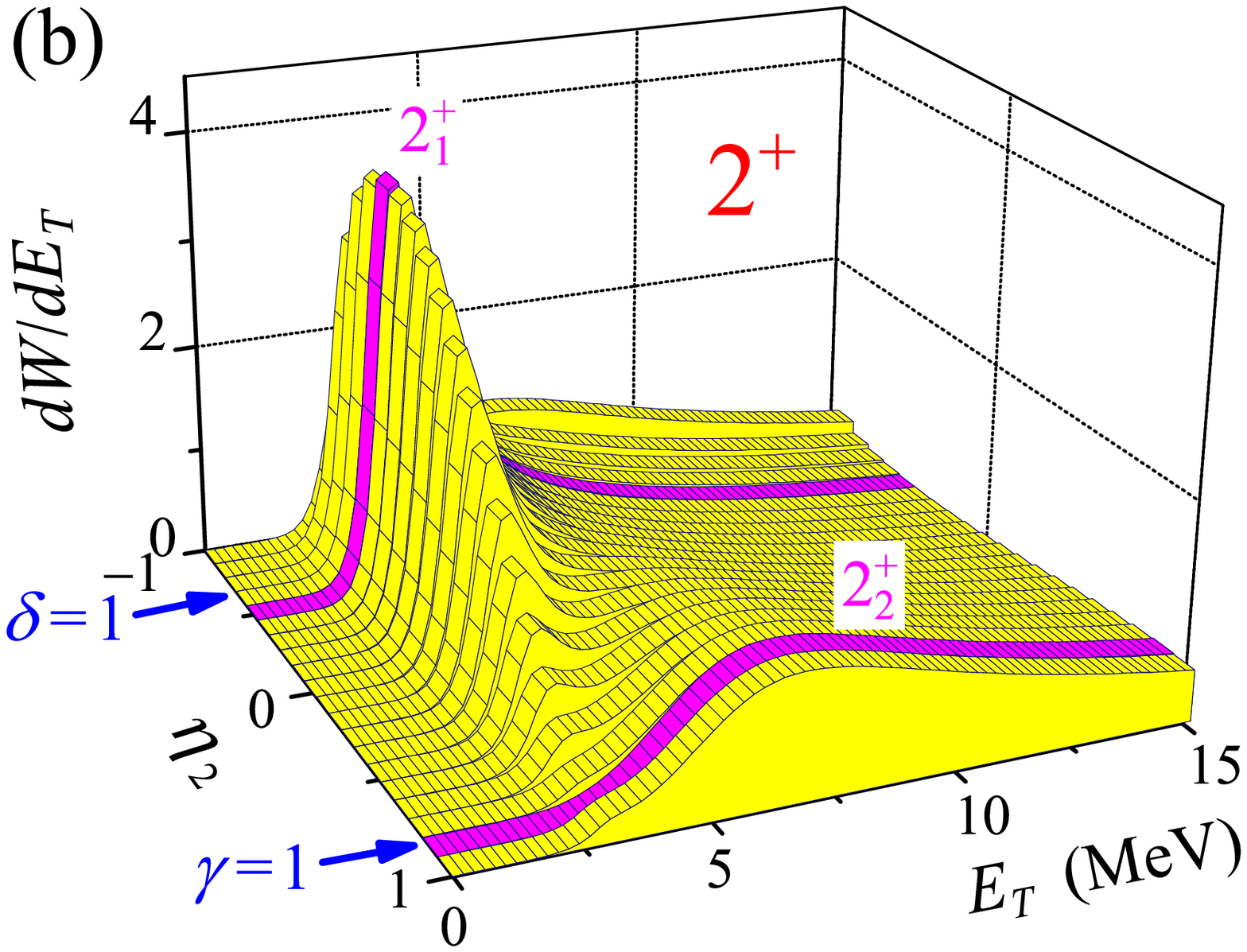} % 44
\caption{Excitation spectra of (a) $0^+$ and (b) $2^+$ (b) states
as function of the spin composition of the source function. The magenta-colored
marked grids correspond to the first and second excitations  $J^+_1$ and
$J^+_2$, see Fig.\ \ref{fig:second}. Arrows show the positions corresponding to
definite shell-model configurations in the source.}
\label{fig:eta-dist}
\end{figure}
%-------------------------------------------------------------------------------

The differential cross section is expressed via the flux induced by the WF
$\Psi^{(+)}$ on the remote surface $S$
\begin{equation}
\frac{d \sigma}{d^3k_{\alpha}d^3k_{p_1}d^3k_{p_2}} \sim  \left. \langle
\Psi^{(+)} | \hat{j} | \Psi^{(+)} \rangle \right|_S.
\label{eq:cross-sect}
\end{equation}
For the reaction considered, this can be rewritten in terms of the
density-matrix formalism
\begin{eqnarray}
\frac{d\sigma}{dq \, dE_T \, d\Omega_5} & = & \textstyle \sum_{JM,J'M'}
\rho_{JM}^{J'M'}(q,E_T) \, \nonumber \\
 & \times & A^{\dagger}_{J'M'}(E_T,\Omega_5) \, A_{JM}(E_T,\Omega_5).
\label{eq:cross-sect-2}
\end{eqnarray}
Some ingredients of the Eqs.\ (\ref{eq:cross-sect},\ref{eq:cross-sect-2}) are
illustrated in Fig.\ \ref{fig:sketch}.  In Eq.\ (\ref{eq:cross-sect-2}), the
contributions of the three-body dynamics (amplitudes $A_{JM}$) and the reaction
mechanism (density matrix $\rho_{JM}^{J'M'}$) are explicitly separated. For
direct reactions, the density matrix has an especially simple form in the frame
with the \textit{z} axis coinciding with the direction of the transferred
momentum $\mathbf{q}$. We use two limiting forms of the density matrix:
\begin{eqnarray}
\rho_{00}^{00} & = & 1 \,, \quad
\rho_{2M}^{2M}= 1/5  \,, \;\;
\rho_{20}^{00}= \rho_{00}^{20}=\cos(\phi_{20})/\sqrt{5},\quad
\label{eq:rhom-nonal} \\
\rho_{00}^{00} & = & 1 \,, \quad
\rho_{20}^{20}= 1  \,, \quad \quad
\rho_{20}^{00}= \rho_{00}^{20}= \cos(\phi_{20}).
\label{eq:rhom-total}
\label{eq:rhomat}
\end{eqnarray}
In the sudden-removal model of Eq.\ (\ref{eq:sour}), there is no alignment of
the final state. This should lead to Eq.\ (\ref{eq:rhom-nonal}) with the
relative phase of the $0^+$ and $2^+$ states $\phi_{20}=0$.
However, it is clear that some alignment should be introduced by a realistic
reaction mechanism. To check our sensitivity to this, we also used the density
matrix of Eq.\ (\ref{eq:rhom-total}) corresponding to the completely aligned
case and also kept $\phi_{20}$ as a parameter in the both above cases. In order
to compare with experiment, Eq.\ (\ref{eq:cross-sect-2}) provides theoretical
input for Monte Carlo (MC) simulations used to deal with the bias introduced by
experimental apparatus.

%-------------------------------------------------------------------------------
\begin{figure}
\includegraphics[width=0.46\textwidth]{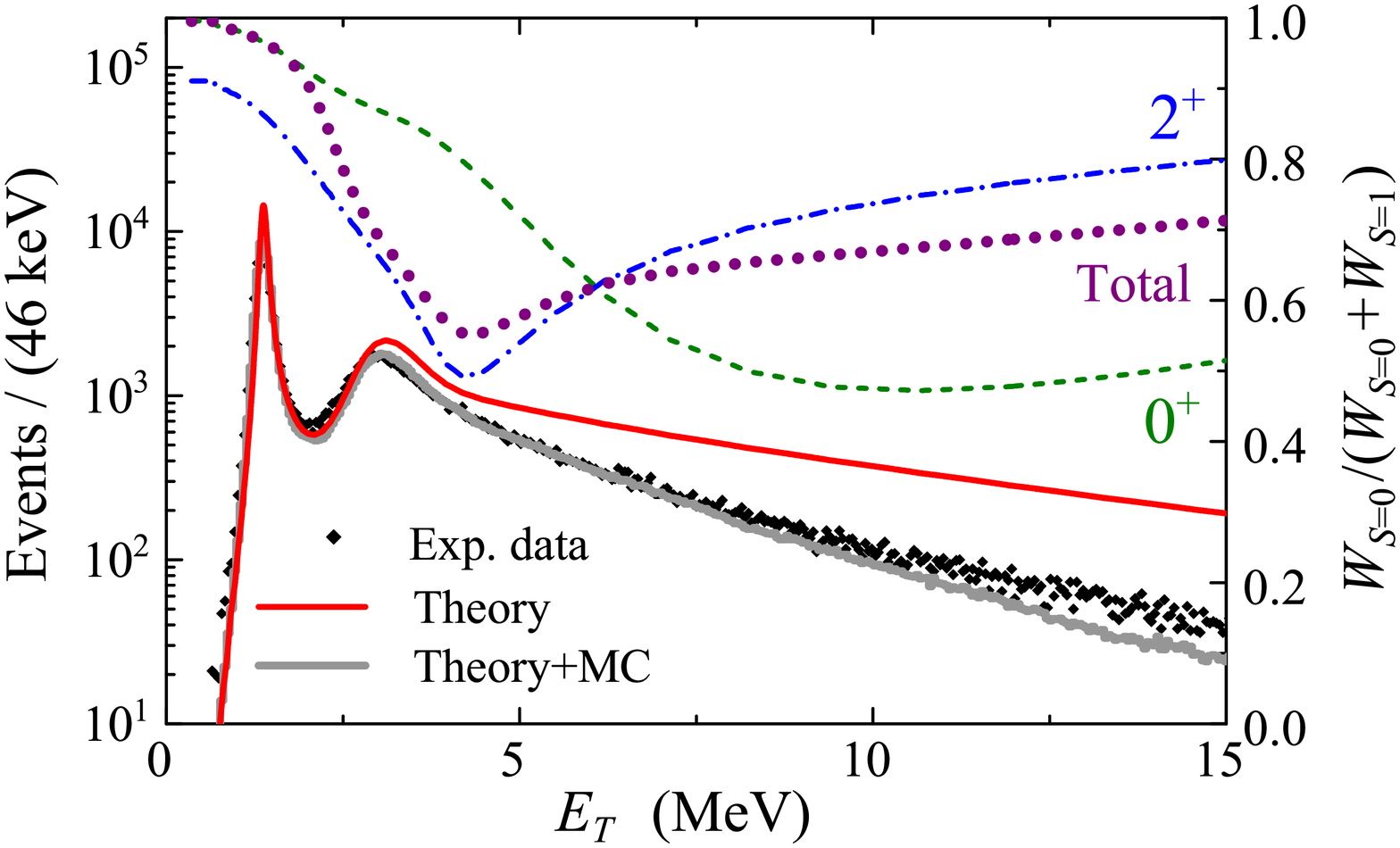}
\caption{Left axis: theoretical $^6$Be excitation
spectrum together with the results of the MC
simulations superimposed on the data \cite{Egorova:2012}.
Right axis: predicted evolution of the spin composition in $^{6}$Be for $0^+$,
$2^+$, and for the
total spectrum fitted to experimental data (the sum of $0^+$ and $2^+$
contributions).}
\label{fig:spec}
\end{figure}
%-------------------------------------------------------------------------------

The reaction model and the initial-state WF can be seen as simplistic. However,
this fits the aim of this paper to provide a ``proof-of-concept'' demonstration,
where interdependencies on different aspects of the model are as transparent as
possible.

%===============================================================================

\textit{Sensitivity of the $^{6}$Be spectrum to $W_{S=0}/W_{S=1}$}
%
%===============================================================================
%
is demonstrated in Fig.\ \ref{fig:eta-dist}.  For some range of parameters, the
conventional picture of $^{6}$Be excitation is
observed with the $0^+$ and  $2^+$  states at 1.37 MeV and 3.05 MeV,
respectively, with significant variations taking place only in the high-energy
``tail'' of the distribution. However, with other variations of the parameters,
the ``normal'' states of $^{6}$Be ``fade'' and even completely disappear, while
new broad peaks arise for these two states at $\sim12$ MeV and $\sim 7$ MeV,
respectively. Thus, our reaction model  links variations of the $^{6}$Be
excitation spectrum to variations of the structure of $^{7}$Be. In this work we
would like to investigate this link quantitatively without reference to
realistic structure of $^{7}$Be in order to find boundaries for possible scale
of effects and establish principal opportunity to use three-body-continuum
spectra as tools to study the spectroscopy of the precursor.

The broad range of spectra obtained for $0^+$ and $2^+$ continuum provide an
opportunity to fit the experimental spectrum of $^{6}$Be. This fit, obtained in
\cite{Egorova:2012}, is shown in Fig.~\ref{fig:spec} together with the MC and
experimental data. It corresponds to the simple case of pure $S=0$ population in
the source function ($\eta_J = 0$) corresponding to a parameter set for
\emph{initial state} of $\{\alpha,\beta,\gamma,\delta\}=\{0.42,0.3,0.49,0.7\}$.
It should be noted that the spin content of the \emph{final state} may have
nothing in common with the spin content of the source function as the spin
quantum number is not conserved by the Hamiltonian (\ref{eq:shred}). The
predicted \emph{final-state} spin composition (Fig.\ \ref{fig:spec}, right
axis), evolves rapidly between the location of the $0^+$ peak and past the
location of the $2^+$ peak indicating  important modifications of the nuclear
structure in this energy region.

%===============================================================================

\begin{figure}
\includegraphics[width=0.42\textwidth]{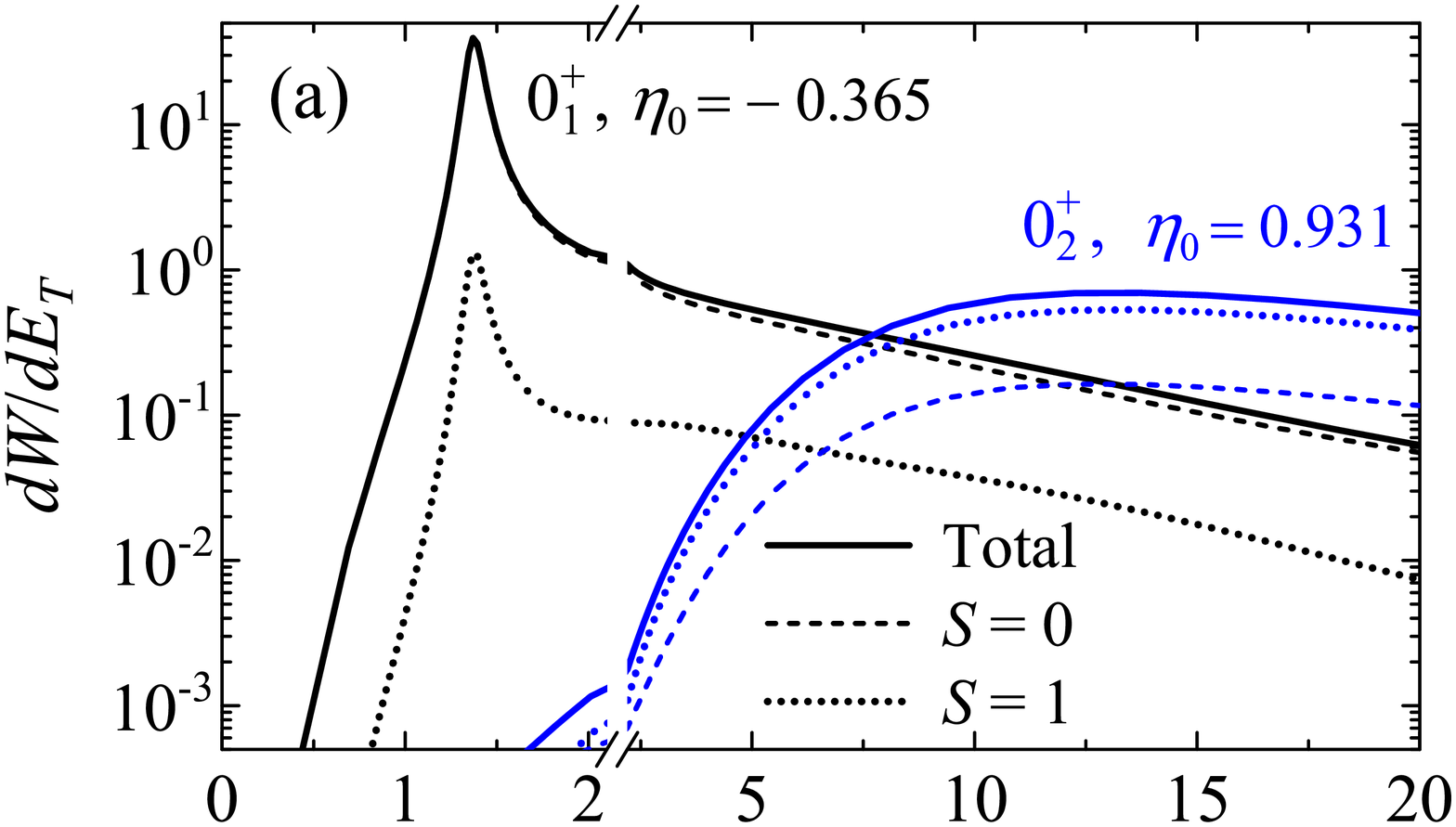} %42
\includegraphics[width=0.42\textwidth]{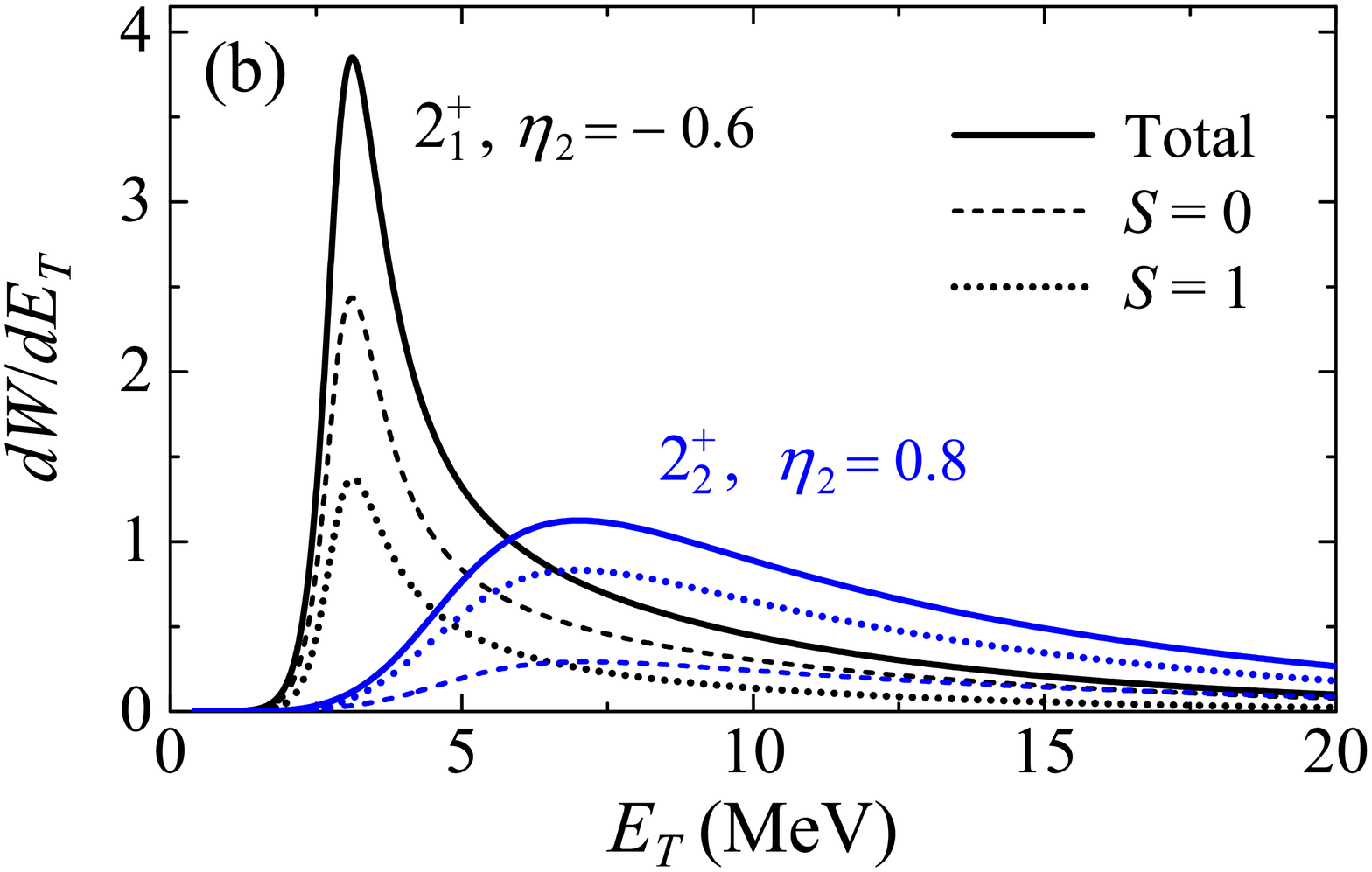}
\caption{(Color online) Excitation spectra and spin composition for
the $^6$Be continuum explicitly representing the
first and second resonances for (a) $J^{\pi}=0^+$ and (b) $2^+$ .
Note, the broken abscissa in (a).}
\label{fig:second}
\end{figure}

\textit{Second $0^+$ and $2^+$ states}
%
%===============================================================================
%
--- The above observations allow us to determine the properties of the $0^+_2$
and $2^+_2$ states. We just need to choose the parameter settings minimising
population of the normal resonant peaks. An important feature of ``pure'' first
and second states, illustrated in Fig.\ \ref{fig:second} (see also arrows in
Fig.\ \ref{fig:eta-dist}), is that the spin composition depends  weakly on
energy. The existence of a common spin structure independent on energy, allows
us to interpret $J^+_1$ and $J^+_2$ as different states, although they can be
represented by (relatively) broad overlapping structures. This property of
``pure'' states is in  sharp contrast with the spin evolution for the
``composite'' situation of Fig.\ \ref{fig:spec}. The spin-content ratios
$0^+_1/0^+_2$ and $2^+_1/2^+_2$ provide a simple structural idea: these states
are partners, in the sense they are the orthogonal combinations of the $S=0$ and
$S=1$ configurations. E.g.\ for $2^+$ states it can be seen that the
$W_{S=0}/W_{S=1}$ ratio is 2:1 for the first and $\sim$1:2 for the second
resonance. It can be seen from Figs.\ \ref{fig:eta-dist} and \ref{fig:second}
that the ``pure'' states are best populated from source functions which are
close to pure shell configurations in Eq.\ (\ref{eq:psi-7be}) and in fact their
structure is reasonably close to such pure shell configurations.

%-------------------------------------------------------------------------------
%-------------------------------------------------------------------------------

The above insight on the structure of the $0^+_2$ and  $2^+_2$ states of
$^{6}$Be provides clear guidelines for experimental searches: the reaction
mechanism should enrich the $S=1$ component in the final state.

%===============================================================================

\begin{figure}
\includegraphics[width=0.42\textwidth]{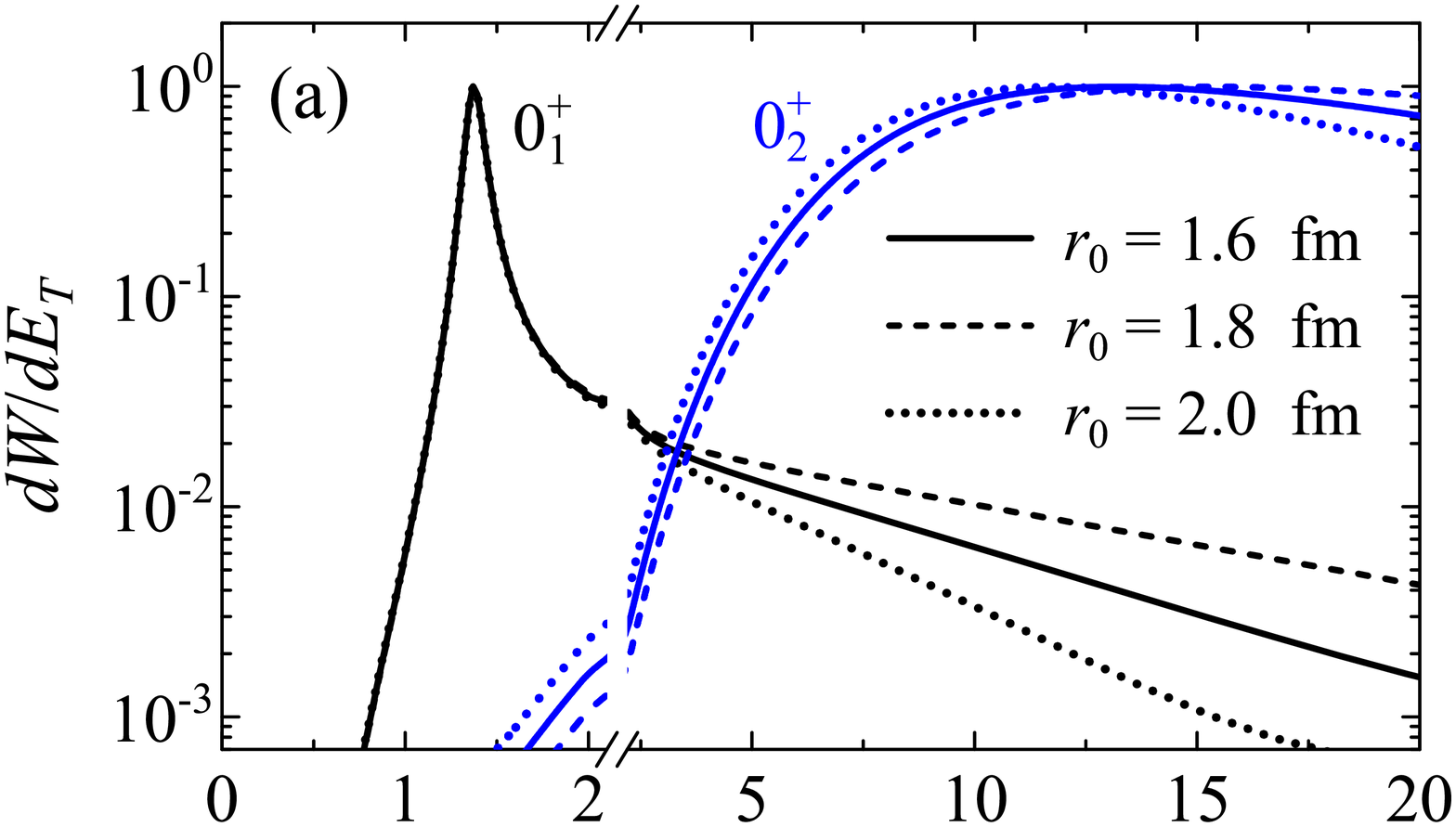}
\includegraphics[width=0.42\textwidth]{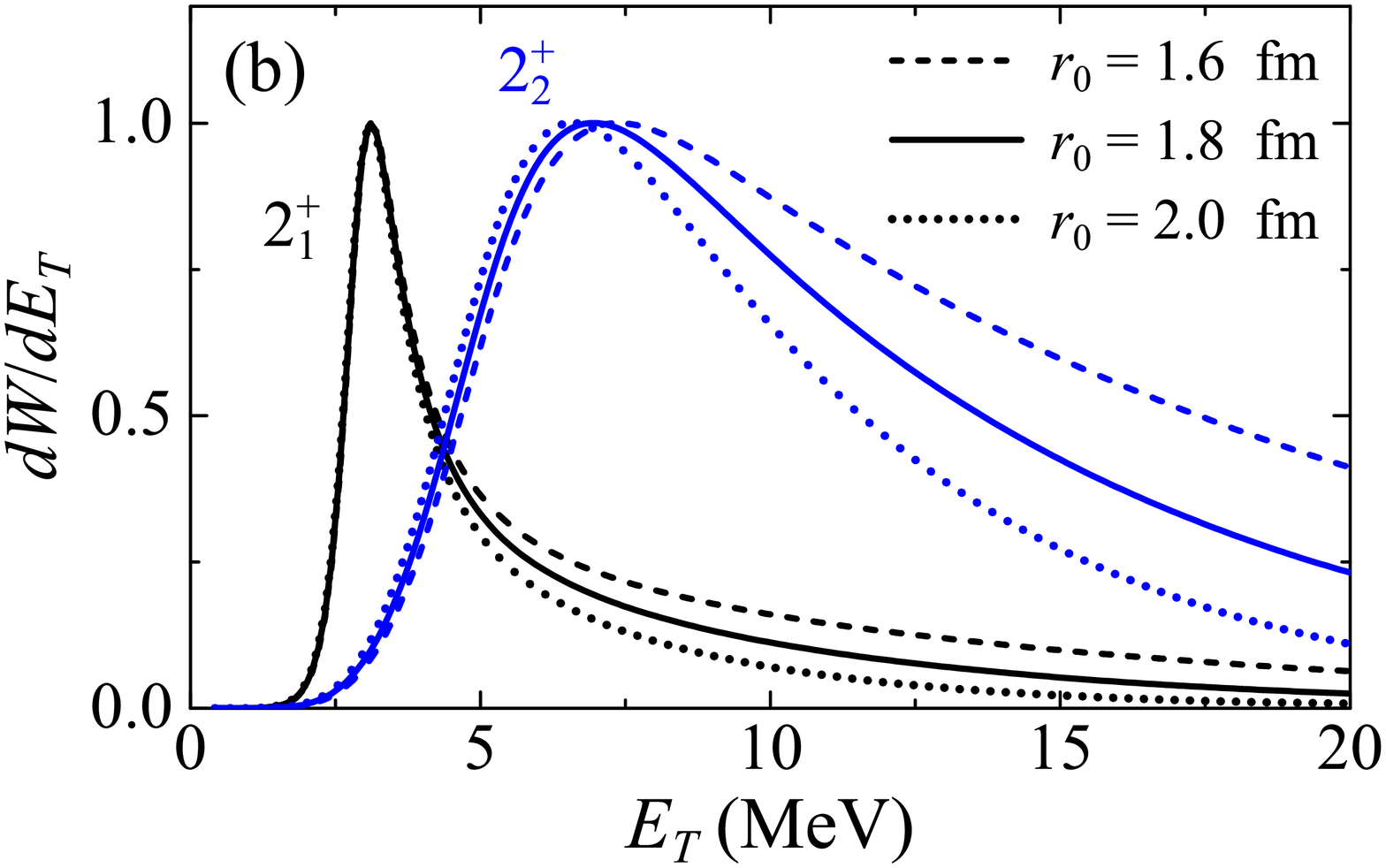}
\caption{(Color online) Dependence of the $^{6}$Be excitation spectra on the
radius parameter $r_0$ of the $^{7}$Be WF. All spectra are normalized to a
maximum value of unity.}
\label{fig:rad-dep}
\end{figure}

\textit{Radial dependence.}
%
%===============================================================================
%
--- The sensitivity to the radial characteristics of the source is presented in
the Fig.\ \ref{fig:rad-dep}. It is practically nonexistent for $0^+_1$:
variations take place only in the ``tail'', a few decay widths higher than the
resonance position. The  $2^+_1$ sensitivity is quite small: the resonance width
is affected on the level of $15\%$. However, with increasing excitation energy
of the states (and hence with increasing decay width) the effect grows. There is
about 0.8 MeV uncertainty of the $2^+_2$ position connected with the radial
extent of the source, and the profile of the cross section is strongly affected.
Variation of $0^+_2$ properties is so large (few MeV in peak position) that
that the properties of such a state cannot be discussed without a detailed
account of the reaction mechanism.

%-------------------------------------------------------------------------------
%-------------------------------------------------------------------------------

%===============================================================================

\textit{Three-body correlations.}
%
%===============================================================================
%
--- The three final momentum vectors in Eq.\ (\ref{eq:cross-sect}) and
Fig.~\ref{fig:sketch} can be transformed to the transferred momentum
$\mathbf{q}$ and the Jacobi momenta $\mathbf{k}_x$, $\mathbf{k}_y$ in Eq.\
(\ref{eq:cross-sect-2}):
\begin{eqnarray}
{\bf k}_x  =  \frac{A_2 {\bf k}_1-A_1 {\bf k}_2 }{A_1+A_2} \, ,  \,\;
{\bf k}_y  =  \frac{A_3 ({\bf k}_1+{\bf k}_2)-(A_1+A_2) {\bf k}_3}
{A_1+A_2+A_3} , \nonumber \\
E_T =E_x+E_y=k^2_x/2M_x + k^2_y/2M_y ,  \qquad \qquad
\label{eq:corel-param}
\end{eqnarray}
where $M_x$ and $M_y$ are the reduced masses of the $X$ and $Y$ subsystems, see
Ref.\ \cite{Grigorenko:2009c} for details.

The orientation of $\mathbf{q}$ is not a dynamical variable of the model
Eq.~(\ref{eq:sour}) and thus is not present in Eq.\ (\ref{eq:cross-sect-2}).
The five-dimensional ``hyperspherical solid angle'' $\Omega_5$ includes two
degrees of freedom describing the ``internal correlations'' of the three-body
system which are ordinarily considered as being completely defined by the
dynamics of the three-body motion itself. The parameters
\begin{equation}
\varepsilon = E_x/E_T \, ,\qquad \cos(\theta_k)=(\mathbf{k}_{x} \cdot
\mathbf{k}_{y}) /(k_x\,k_y) \, , \nonumber \\
\label{eq:ener-ang}
\end{equation}
provide what we call ``complete energy-angular correlations''. For $k_3
\rightarrow k_{\alpha}$ we get the ``T'' Jacobi system, where $\varepsilon$
describes the energy correlation in the $p$-$p$ channel. For $k_3 \rightarrow
k_{p_i}$, the correlations are obtained in one of two possible ($i=1,2$)
``Y'' Jacobi systems, where $\varepsilon$ describes the energy correlation in
the $\alpha$-$p$ channel.

The other three degrees of freedom (Euler angles) define ``external
correlations'' as they describe the orientation of the three-body system as a
whole. Correlations for the ``external'' degrees of freedom are evidently
defined by the reaction mechanism.

%===============================================================================

\textit{Alignment and interference.}
%
%===============================================================================
%
--- The internal three-body correlations for excited states cannot be separated
from external information in any practical experiment. The excited states
typically have nonzero $J$  making alignment possible and they reside on the
``tails'' of the lower-energy excitations so that their amplitudes can
interfere. From a theoretical point of view, the inclusive energy $\varepsilon$
distributions should be free of interference effects and experience tell us that
the angular $\cos(\theta_k)$ distributions are only weakly affected. However in
 experiments, the bias of the apparatus introduces cut-offs and distortions
which may induce correlations via loss of orthogonality for configurations with
different angular momenta. Such ``induced correlations'' are specific for the
experiment and should be evaluated by careful MC studies.

\begin{figure}
\includegraphics[width=0.48\textwidth]{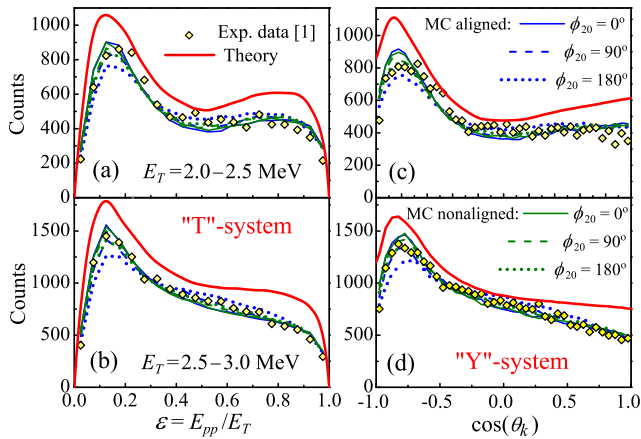}
\caption{(a,b) ``Internal'' energy distributions in the Jacobi
``T'' system and (c,d) angular distributions in the Jacobi ``Y'' system. Upper
and lower rows of panels correspond to different $E_T$ ranges. The experimental
data are shown by the hollow diamonds. Theoretical curve (thick, solid, red)
is given in each panel with some offset to simplify a perception. MC curves for
different
alignment/interference settings are explained in the legend.}
\label{fig:align-interf}
\end{figure}

A demonstration of alignment/interference effects on  experimental results is
provided in Fig.\ \ref{fig:align-interf}. As it is clear that the effects are
largest for strong $0^+/2^+$ mixing, we have chosen the $E_T=2-3$ MeV energy
range of the data \cite{Egorova:2012}. MC results for the two limiting cases of
completely-aligned (\ref{eq:rhom-total}) and ``isotropic'' (\ref{eq:rhom-nonal})
density matrices, each with three different phase settings $\phi_{20}=\{
0,\pi/2,\pi \}$,  are shown. In Ref.~\cite{Egorova:2012}, we used the settings
of Eq.~(\ref{eq:rhom-nonal}) with $\phi_{20}=0$ which is consistent with the
data and corresponds to the model of Eq.~(\ref{eq:sour}). However, at the moment
we cannot exclude that reality is different: further analysis of the data is
needed. Fig.\ \ref{fig:align-interf} shows the energy distribution in the Jacobi
``T'' system and the angular distribution in the Jacobi ``Y'' system which
appear to be the most sensitive of the correlation observables. The scale of
local variations in the MC distributions is about $20\%$. The curves are well
separated and some deviate considerably from the experimental results.

%-------------------------------------------------------------------------------
%-------------------------------------------------------------------------------

It should be understood that all the MC curves correspond to the \emph{same
theoretical distribution} and their variations are due to the \emph{bias
introduced by experimental setup}. Thus we conclude that in analysing
high-precision correlation data for  excited states where interference/alignment
effects become possible, a consistent treatment of the reaction mechanism
becomes inevitable. The effect depends strongly on the quality of the
experimental setup; it should vanish for an ``ideal'' instrument. The moderate
level of variations in Fig.\ \ref{fig:align-interf} is connected with the very
high efficiency of the setup in Ref.\ \cite{Egorova:2012}. On lower-quality
setups, alignment/interference effects can produce very large and poorly
controlled modifications. This could be a part of the explanation for the strong
deviation of the data of Ref.~\cite{Papka:2010} from the other recent
experimental studies \cite{Grigorenko:2009c,Fomichev:2012,Egorova:2012}.

%===============================================================================

\textit{On existence of higher-energy negative parity excitations.}
%
%===============================================================================
%
--- The availability of higher excitations in the spectrum of $^{6}$Be
\cite{Egorova:2012}  may  drastically affect the proposed interpretation of the
data. In Ref.~\cite{Fomichev:2012},  a strong population of presumably
$\{0^-,1^-,2^-\}$ states in the charge-exchange reaction induced by $\Delta L=1$
transitions was observed above the first $2^+$ state. We were cautious about the
population of such negative-parity states in the data of
Ref.~\cite{Egorova:2012}. A dedicated search was performed for asymmetries in
certain distributions due to the interference of positive and negative states.
However, no significant indication of these asymmetries was found. Also, the
dominating shell structure of the $^{7}$Be precursor does not imply a strong
population of negative-parity states in $^{6}$Be within the sudden-removal
model. Thus we have confined our interpretation of the $^{6}$Be continuum with
just the $0^+$ and $2^+$ states.

%===============================================================================

\textit{Previous theoretical results on $^{6}$Be.}
%
%===============================================================================
%
--- Three-cluster calculations (microscopic and three-body) for $^{6}$Be have
been performed in a number of studies
\cite{Danilin:1993,Csoto:1994,Vasilevsky:2001,Descouvemont:2006,Garrido:2007,%
Michel:2010}. In Ref.~\cite{Vasilevsky:2001}, the energies and widths
$\{E_R,\Gamma\}$ of the $0^+_2$ and $2^+_2$ resonances were predicted to be
$\{3.5,6.1\}$ and $\{5.2,5.6\}$ MeV. These values are quite different from our
predictions $\{\sim 12,\sim 14\}$ and $\{\sim 7,\sim 7\}$ MeV. The reason for
this discrepancy is easily understood: our prediction is that the second $0^+_2$
and $2^+_2$ states are ``spin-complements'' of the first $0^+_1$ and $2^+_1$
states. States of such a nature could not have been obtained in the calculations
of Ref.~\cite{Vasilevsky:2001} as the $S=1$ component was omitted in their model
space.

\begin{figure}[tb]
\begin{center}
\includegraphics[width=0.49\textwidth]{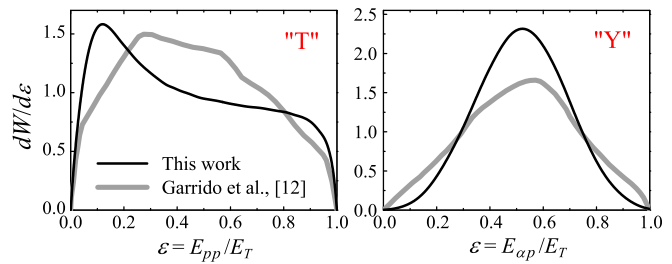}
\end{center}
\caption{Comparison between the energy distributions for the $2^+$ resonance of
$^{6}$Be in ``T'' and ``Y'' Jacobi systems obtained in the present work and from
Ref.~\cite{Garrido:2007}.}
\label{fig:comp}
\end{figure}

Among these other theoretical studies, only in Ref.~\cite{Garrido:2007} are
theoretical correlations (energy distributions for $2^+$) presented. The
predicted correlations for the $0^+$ and $2^+$ states (presumably related to
\cite{Garrido:2007}) are provided in \cite{Papka:2010}, but in a form which make
them difficult to interpret as only MC results specific for experimental setup
of that work are shown. The distributions of Ref.~\cite{Garrido:2007} are not in
agreement with our results (see Fig.\ \ref{fig:comp}) and hence with the recent
highly-accurate data of Refs.~\cite{Fomichev:2012,Egorova:2012}. This
observation sheds doubts on applicability of the methods of \cite{Garrido:2007}
to Coulombic three-body decays in general. Our work imposes a new standard of
a sophistication required from theoretical calculations to analyse  modern
high-precision data including correlations.

%-------------------------------------------------------------------------------
%-------------------------------------------------------------------------------

%===============================================================================

\textit{Conclusions.}
%
%===============================================================================
%
--- This work provides important qualitative insights into the question of which
aspects of three-body decays can be understood based on the dynamical
description of the final state alone and which also require an adequate
treatment of the initial state and the reaction mechanism. The major and novel
observations of this work are the following.

\noindent (i)  The population of the $^{6}$Be $0^+_1$ ground state is very
stable to variations of the initial-state structure and the reaction mechanism.
However, the latter effects become increasingly important with increasing
excitation. No sensible description of the continuum above $5-7$ MeV can be
given without their proper treatment.

\noindent (ii) The excitation spectrum of the three-body continuum of $^{6}$Be
up to $15-20$ MeV
is found to be very sensitive to the spin composition of the
source function in Eq.\ (\ref{eq:shred}). In our calculations, this was
parametrized in terms of the spin content of precursor. Thus we find that
three-body decay can be used as a sensitive instrument of nuclear spectroscopy
if the reaction mechanism is well established.

\noindent (iii)  A procedure to identify the second $0^+_2$ and $2^+_2$ states
of $^{6}$Be is proposed. These states are found to be ``spin-complements'' of
the well-known $0^+_1$ and $2^+_1$ states, providing a guideline for
their experimental observation.

\noindent (iv)  Alignment and interference effects are observable in
experimental data due to the unavoidable experimental bias. They were found to
have an important impact on the measured three-body correlations.
Proton-proton energy distributions are especially affected.
This indicates that caution is needed in studies of $N$-$N$ correlations in
decays of excited (or/and broad) three-body states in general.

%===============================================================================
%
\textit{Acknowledgments.}
%
%===============================================================================
%
--- L.V.G.\ and I.A.E.\ are supported by the Helmholtz Association under grant
agreement IK-RU-002 via FAIR-Russia Research Center. L.V.G.\ is supported by
Russian Foundation for Basic Research 11-02-00657-a and Ministry of Education
and Science NS-215.2012.2 grants.  R.J.C.\ is supported by the U.S.\ Department
of Energy, Division of Nuclear Physics under grants DE-FG02-87ER-40316.

%###############################################################################

%###############################################################################

\end{document}